\documentclass[prd,showkeys,twocolumn]{revtex4}
\usepackage{graphicx}

\def\gap{\;\rlap{\lower 2.5pt
 \hbox{$\sim$}}\raise 1.5pt\hbox{$>$}\;}
\def\lap{\;\rlap{\lower 2.5pt
   \hbox{$\sim$}}\raise 1.5pt\hbox{$<$}\;}
\def\gsim{\;\rlap{\lower 2.5pt
 \hbox{$\sim$}}\raise 1.5pt\hbox{$>$}\;}
\def\lsim{\;\rlap{\lower 2.5pt
   \hbox{$\sim$}}\raise 1.5pt\hbox{$<$}\;}

\def\spose#1{\hbox to 0pt{#1\hss}}
\def\lta{\mathrel{\spose{\lower 3pt\hbox{$\mathchar''218$}}
     \raise 2.0pt\hbox{$\mathchar''13C$}}}
\def\gta{\mathrel{\spose{\lower 3pt\hbox{$\mathchar''218$}}
     \raise 2.0pt\hbox{$\mathchar''13E$}}}
\newcommand{\beq}{\begin{equation}}
\newcommand{\eeq}{\end{equation}}
\newcommand{\be}{\begin{equation}}
\newcommand{\ee}{\end{equation}}
\newcommand{\nn}{\nonumber \\}
\newcommand{\ls}{\mathrel{\raise1.16pt\hbox{$<$}\kern-7.0pt 
\lower3.06pt\hbox{{$\scriptstyle \sim$}}}}         
\newcommand{\gs}{\mathrel{\raise1.16pt\hbox{$>$}\kern-7.0pt 
\lower3.06pt\hbox{{$\scriptstyle \sim$}}}}         

\long\def\comment#1{}

\def\fun#1#2{\lower3.6pt\vbox{\baselineskip0pt\lineskip.9pt
  \ialign{$\mathsurround=0pt#1\hfil##\hfil$\crcr#2\crcr\sim\crcr}}}
\def\lap{\mathrel{\mathpalette\fun <}}
\def\gap{\mathrel{\mathpalette\fun >}}
\newcommand{\ba}{\begin{eqnarray}}
\newcommand{\ea}{\end{eqnarray}}
\newcommand{\vk}{{\bf{k}}}

\newcommand{\hgamma}{\hat{\gamma}}

\begin{document}
\bibliographystyle{apsrev.bst}
\title{Dark energy and cosmic microwave background bispectrum}
\author{Licia Verde}
\altaffiliation{Dept of Physics \& Astronomy, Rutgers University, 136 Frelinghuysen Road, Piscataway NJ 08854-8019, USA}
\author{David N. Spergel}
\altaffiliation{Institute for Advanced Study, Princeton,  NJ 08540, USA}
\affiliation{Depart. of Astrophysical Sciences, Peyton
Hall, Princeton University, Ivy Lane, Princeton, NJ 08544 1001, USA}
\email{lverde,dns@astro.princeton.edu}

\begin{abstract}
We compute the cosmic microwave background  bispectrum arising from the cross-correlation of primordial, lensing and Rees-Sciama signals. 
The amplitude of the bispectrum signal is 
sensitive to gthe matter density parameter,
$\Omega_0$, and the equation of state of the dark energy, which
we parameterize by $w_Q$.
We conclude that the dataset of the Atacama Cosmology Telescope, combined with MAP 2-year data or the
Planck data set alone will allow us to break the degeneracy
between $\Omega_0$ and $w_Q$   that arises from 
the analysis of CMB power
spectrum. In particular a joint measurement of  $\Omega_0$ and $w_Q$
with 10\% and 30\% error on the two parameters respectively, at the 90\% confidence level can
realistically be achieved.   
\end{abstract}
\maketitle
\section{Introduction}
Recent results suggest that the Universe is flat and dominated by a negative
pressure (dark
energy) component (e.g., [1] and references therein) which can be characterized by its equation of state
parameter $p/\rho=w_Q$. A constant vacuum energy (i.e. a cosmological constant)
has $w_Q=-1$, while a generic $w_Q<0$ dark energy component is refereed to as
``quintessence'' (e.g., [2]).
Analysis of the power spectrum alone of 
forthcoming Cosmic Microwave Background (CMB) experiments will still present some 
degeneracy between $w_Q$ and $\Omega_0$ (after marginalization over other
cosmological parameters)(e.g., [3]).   

Here we consider the constraints on $w_Q$-$\Omega_0$ that can be obtained from
the analysis of the small scale CMB bispectrum for two experimental settings: the
combination of MAP [4] and Atacama Cosmology Telescope (ACT; [5]), and the
Planck surveyor  satellite [6,7]. 

Under the assumption of Gaussian initial conditions, the cross-correlation
of  primordial, gravitational lensing and Rees-Sciama (RS) [8] signals is
the dominant contribution to the CMB bispectrum after that of the
Sunyaev-Zeldovich (SZ) effect [9] and of point sources. 
Since each term in the bispectrum has a different function form,
we can separate these signals without major loss of information [e.g., 10].

This paper is organized as follows.
In section 2 we present the expression of the primordial-lensing-RS
cross-correlation bispectrum. In section 3 we forecast the error
on the joint determination of $\Omega_0$ and $w_Q$ from the bispectrum
analysis. In section 4 we conclude that it is possible  to break the
degeneracy between $\Omega_0$ and $w_Q$ that arises from the CMB power
spectrum analysis alone. 

\section{The primordial-lensing-Rees-Sciama cross correlation bispectrum}

We wish to compute the effect on the CMB bispectrum of the coupling between the RS and
the weak lensing. The weak lensing effects allow us to probe the low redshift
Universe through the lensing imprint that foreground structures leave on the
primordial CMB signal. 

Neglecting galactic contamination, the CMB temperature at any position in the sky $\hgamma$ can be expanded as:
\ba
\frac{\Delta T(\hgamma)}{T}&\simeq& \frac{\Delta T^P(\hgamma)}{T}+\nabla
\frac{\Delta T^P(\hgamma)}{T}\cdot
\nabla\Theta(\hgamma)\nn
& +&\frac{\Delta T^{NL}(\hgamma)}{T}+\frac{\Delta T^{SZ}(\hgamma)}{T}+\ldots
\label{eq:deltattotal}
\ea
where the first term is the primordial signal, the second term is the
gravitational lensing effect, the third term is the RS contribution and the
last term denotes the SZ effect. The frequency dependence of the
SZ term makes it possible to separate out this contribution and we will
therefore ignore this term in what follows. Point sources contribution to the
bispectrum signal can be separated out without major loss of information [10]. Other contributions to
$\Delta T/T$ such as e.g., the Ostriker-Vishniac effect [11] will have zero or
sub-dominant contribution to the CMB bispectrum. 

The primordial contribution can be expressed as:
\be
\frac{\Delta T^P(\hgamma)}{T}=\int\frac{d^3k}{(2\pi)^3}\exp(i \vk\cdot \hgamma
r_*)\tilde{\Phi}(k) g(k)
\ee
where $g$ denotes the radiation transfer functional, $\tilde{\Phi}$ denotes the
Fourier transform of the gravitational potential perturbation ${\Phi}$, and $r_*$ denotes the conformal distance to the last scattering surface.
The lensing potential $\Theta(\hgamma)$ is the  projection of the gravitational
potential:
\be
\Theta(\hgamma)=-2\int_0^{r_*}dr\frac{r_*-r}{rr_*}\Phi(r,\hgamma r).
\ee  

The RS effect arises from  a combination of two effects: the late-time decay of gravitational potential fluctuations in a non-Einstein-de Sitter
Universe ---strictly called Integrated Sachs-Wolfe effect [12]--- and the non-linear
growth of density fluctuations [8] along the photon path. The third term
in Eq. (\ref{eq:deltattotal}) can be expressed as
\be
\frac{\Delta T^{NL}(\hgamma)}{T}=2\int dr  \frac{\partial}{\partial
t}\Phi^{NL}(r, \hgamma r).
\ee

The CMB bispectrum is defined as:
\be
B_{\ell_1 \ell_2 \ell_3}^{m_1 m_2 m_3}=\langle a_{\ell_1}^{m_1}a_{\ell_2}^{m_2}a_{\ell_3}^{m_3}\rangle= \left(^{\ell_1\;\;\;\ell_2\;\;\;\ell_3}_{m_1m_2m_3} \right)B_{\ell_1 \ell_2\ell_3} 
\ee
where $a_{\ell}^m$  are the coefficients of the spherical harmonics expansion
of  the observed CMB temperature  fluctuation: 
\be
a_{\ell}^m=\int d^2\hgamma \frac{\Delta T(\hgamma)}{T} Y^{*m}_l(\hgamma)\;,
\label{eq:sphharmtr}
\ee 
$\left(^{\ell_1\;\;\;\ell_2\;\;\;\ell_3}_{m_1m_2m_3} \right)$ is the
Wigner three-J symbol, and the last equality results from symmetry reasons
(e.g., [13,14]).

By applying Eq.(\ref{eq:sphharmtr}) to Eq.(\ref{eq:deltattotal}) we obtain
(cf. [15,10]),
\ba
a_{\ell m}& = & a_{\ell}^{mP}+\sum_{\ell'\ell''m' m''}(-1)^{m+m'+m''}{\cal H}^{-m m' m''}_{\ell
\ell' \ell''} \nn
& \times &\frac{\ell'(\ell'+1)-\ell(\ell+1)+\ell''(\ell''+1)}{2}a^{m'P*}_{\ell'}\Theta^{*
-m''}_{\ell''}\nn
& + &a^{m NL}_{\ell} 
\ea
where ${\cal H}$ denotes Gaunt integral
\ba
{\cal H}^{m_1 m_2
m_3}_{\ell_1\ell_2\ell_3} &=&\sqrt{\frac{(2\ell_1+1)(2\ell_2+1)(2\ell_3+1)}{4\pi}} \nn
&\times & \left(^{\ell_1\;\;\;\ell_2\;\;\;\ell_3}_{0\;\;\;\;0\;\;\;\;0}\right)
\left(^{\ell_1\;\;\;\ell_2\;\;\;\ell_3}_{m_1\;
m_2\; m_3}\right)
\ea
and for the bispectrum we obtain
\ba
B_{\ell_1 \ell_2 \ell_3}^{m_1 m_2 m_3}& = &{\cal H}^{m1 m2
m3}_{\ell_1\ell_2\ell_3}\frac{\ell_1(\ell_1+1)-\ell_2(\ell_2+1)+\ell_3(\ell_3+1)}{2}\nn
& \times & C_{\ell_1}^P\langle\Theta^{*m_3}_{\ell_3}a_{\ell3}^{NL
m3} \rangle + 5 \mbox{ permutations}\;.
\ea

Following the steps outlined in the Appendix we
obtain an expression for 
${\cal Q}(\ell_3)\equiv \langle \Theta_{\ell_3}^{*m_3} a_{\ell3}^{NL
m3}\rangle$ (cf. [15,16]) ,
\ba
{\cal Q}(\ell_3)\simeq2\int_0^{z_{*}}\!\!\frac{r(z_{*})-r(z)}{r(z_{*})r(z)^3}\left.\frac{\partial}{\partial
z}P_{\Phi}(k,z)\right|_{k=\frac{\ell_3}{r(z)}}\!\!\!dz
\label{eq:smallangleql}
\ea 
where $z_{*}$ denotes the redshift of the last scattering surface and 
$P_{\Phi}(k,z)$ denotes the power spectrum of the gravitational potential
at redshift $z$; it has to be evaluated at $k=\ell_3/r$ and then derived with
respect to $z$.

Since we assume a flat Universe, $r(z)$, the conformal distance from the observer at $z=0$ is 
\be
r(z)=\frac{c}{H_0}\int_0^z\frac{dz'}{E(z')}
\ee
where
\be
E(z')=\sqrt{\sum_i\Omega_i(1+z')^{3(1+w_i)}}
\ee
and $\Omega_i$ are the normalized densities of the various energy components of
the Universe. The exponents depend on how each component density varies with
the expansion of the Universe, $\rho_i\propto a^{n_i}$, where  $a$ is the scale
factor and $n_i=3(1+w_i)$, and consequently $w_i$ is the equation of
state parameter for the $i$ component. For example for  matter $n=3$, for a
cosmological constant $w=-1, n=0$.

\subsection{Computation of ${\cal Q}(\ell)$}
We compute ${\cal Q}(\ell)$ numerically for different combinations of $\Omega_0$ and
$w_Q$ for COBE-normalized models. We assume a flat Universe and set $\Omega
h^2=0.17$. This is justified by the fact that, from MAP 2-year data, this
quantity should be known to better than 5\% accuracy [17]. 

The gravitational potential power spectrum at any given redshift $z$ is related to the density power
spectrum ($P$) via:
\be
P_{\Phi}(k,z)=\left(\frac{3}{2}\Omega_0\right)^2\left(\frac{H_0}{k}\right)^4 P(k,z) (1+z)^2\;.
\ee
In the linear regime
\be
P^{LIN}(k,z)=Ak^{n_s}T(k)^2\left(\frac{g(z)}{g(z_{*})}\frac{(1+z_{*})}{(1+z)}\right)^2
\label{eq:Plin}
\ee
where $T(k)$ denotes the matter transfer function, $g(z)$ the correction to
the growth factor due to the presence of dark energy, $A$ is the amplitude of
the primordial power spectrum (see [18]), and $n_s$ denotes the primordial spectral slope.

In the case of  $w_Q=-1$, (i.e. for a cosmological constant)  the expression for $g(z)$ is:
\be
g(z)=\frac{5}{2}\frac{\Omega_z}{\Omega_z^{4/7}-\Lambda_z+(1+\Omega_z/2)(1+\Lambda_z/70)}\;,
\ee
otherwise there is a correction factor [19] $(-w_Q)^t$, where 
\ba
t & = & -(0.255+0.305w_Q+0.0027/w_Q)(1-\Omega_z)\nn
& - & (0.366+0.266w_Q-0.07/w_Q)\ln\Omega_z.
\ea
Here $\Omega_z$ is $
\Omega_z=\Omega_0/\left[\Omega_0+(1-\Omega_0) a^{-3w_Q}\right]$.   
We approximate the  transfer function  with that of Sugiyama [20]. Any corrections for $w\ne-1$ affect only very large scales, that do not
contribute to the signal we are modeling.

Since the signal for ${\cal Q}$ is mostly coming from non-linear scales,
Eq. (\ref{eq:Plin}) is not a good approximation of the actual power spectrum. The nonlinear mass power spectrum can
be obtained for the linear one with the mapping of Peacock and Dodds [21] for $w_Q=-1$
while to generalize the mapping to $w_Q\ne-1$ we use the expression of
Ma et al. [18]:
\be
\Delta^{2\; NL}(k)=G\left(\frac{\Delta^{2\;LIN}(k)}{g_{\Delta}^{3/2}\sigma_8^{\beta}}\right)\Delta^{2\;LIN}(k_L)\;,
\ee
where $k_L=k/(1+\Delta^{2\;NL})^{1/3}$, $\Delta^2=2/\pi^2k^3P(k)$,
\be
G(x)=(1+\ln(1+x/2))\frac{1+0.02x^4+c_1x^8/g^3}{1+c_2x^{7.5}}\;,
\ee
$c_1=1.08\times 10^{-4}$, $c_2=2.1\times 10^{-5}$, and  $g_{\Delta}=|w|^{1.3|w|-0.76}g(0)$.

\begin{figure}
\includegraphics[width=8cm]{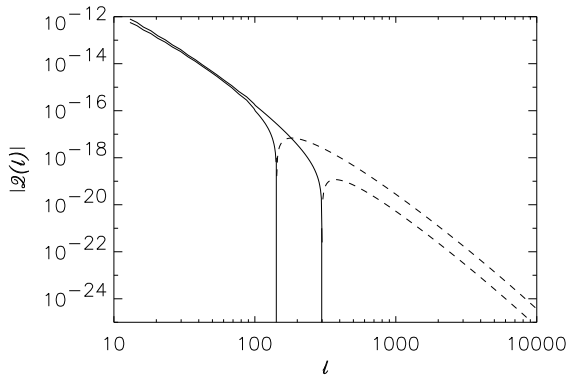}
\caption{Absolute value of ${\cal Q}(\ell)$ for two different
cosmologies: $\Omega_0=0.3, w=-1.0$ (thick line) and $\Omega_0=0.2,
w=-0.2$ (thin line). The solid line indicates ${\cal Q}>0$ while
dashed line indicates ${\cal Q}<0$. If linear theory was a good
approximation for the evolution of the power spectrum, ${\cal Q} \ge
0$: non-linear effects can be important at $\ell \sim 200$.}
\label{fig:ql}
\end{figure}

The sign of $\partial P_{\Phi}/\partial z$ in the integrand of Eq.
(\ref{eq:smallangleql}) is  determined by the balance of two competing contributions:
the decaying of the gravitational potential fluctuations as $z \longrightarrow
0$  and the amplification due to non-linear growth. Both of these are 
sensitive to the cosmological parameters, we thus expect
${\cal Q}(\ell)$ to be sensitive to $w_Q$ and $\Omega_0$.
In Fig. (\ref{fig:ql}) we show the effect of non-linear evolution on  ${\cal
Q}(\ell)$.
The solid line indicates  ${\cal Q}>0$ while dashed line indicates
${\cal Q}<0$. If linear theory was a good approximation for the evolution of
the power spectrum, ${\cal Q} \ge 0$: non-linear effects can be important
at $\ell \sim 200$.

\section{A priori error estimation}
We now wish to evaluate how well forthcoming CMB experiments will be able to
measure $\Omega_0$ and $w_Q$, if we consider the information enclosed in the
primordial-lensing-RS bispectrum in addition to the power spectrum. We will thus estimate the $\chi^2$ on a grid
of cosmological models (cf. [10,15]):
\be
\chi^2=\sum_{\ell_1\ell_2\ell_3}\frac{B_{\ell_1\ell_2\ell_3}^2}{C_{\ell_1}C_{\ell_2}C_{\ell_3}N(\ell_1\ell_2\ell_3)}\;,
\label{eq:chisq}
\ee
where for symmetry we consider $\ell_1\le \ell_2\le \ell_3$;  $N=1$ if
all $\ell$'s are different, $N=2$ if two $\ell$'s are repeated and $N=6$ if
all $\ell$'s are equal.
In deriving Eq.(\ref{eq:chisq}) we have used the fact that the sum over $m$ of the
square of the Wigner three-J symbols is unity.

Following the statistic introduced by [15] the confidence region for $\Omega_0$ and $w_Q$ jointly at the 68.3\%, 90\% and
99\%, is then assumed  to be  given by $\delta\chi^2\equiv
|\chi^2-\chi^2_{\Omega_0,w_Q}|$=2.3, 4.61 and 9.21 respectively.
Of course, the relation between $\Delta\chi^2$ and confidence levels is
strictly correct only if our ``data'' $B_{\ell_1\ell_2\ell_3}$ are normally distributed. 
The central limit theorem ensures that for a large number of
independent data the distribution should asymptotically tend towards a
Gaussian, nevertheless this assumption will need to be tested {\it a
posteriori} or a maximum likelihood technique will need to be used.

In computing $\chi^2$ we make several approximations: first of all the
 expression for ${\cal Q}$ (Eq.\ref{eq:smallangleql}) uses the small
 angle approximation. For the purposes of error estimation this
 approximation is good for $\ell\gap 10$. However most of the signal comes
 from the coupling of very large scales (small $\ell$) to very small scales (large
 $\ell$). Since the exact expression is computationally expensive, we consider
 only $\ell >12$ in our $\chi^2$ calculation on the grid of cosmological
 models. For the standard $\Lambda$CDM model, using the exact expression for
 ${\cal Q}$ at small $\ell$'s, we obtain that the $\chi^2$ is amplified by a
 factor 2 if  $2 <\ell <12$ are also included. 
We can thus infer that this approximately applies to all $\Omega_0$, $w_Q$
 combinations ad that the  cosmological parameters determination can be
 improved consequently, if we include in our analysis all $\ell >2$.

In Eqs. (\ref{eq:smallangleql}) and  (\ref{eq:chisq}) we approximate the
power spectrum ($C_{\ell}$) as composed by three contributions: primordial
($C^P_{\ell}$), Ostriker-Vishniac ($C^{OV}_{\ell}$) and noise ($C^N_{\ell}$). 
$C_{\ell}^P$ is obtained using
CMBFAST [22] up to $\ell=1500$ (with parameters $\Omega_0=0.3$, $\Lambda=0.7$,
$n_s=1$, and, conservatively {\footnote{Other more realistic choices for $\tau$
slightly improve the signal to noise ratio. We thus conservatively set
$\tau=0$ in the error calculation.}}, $\tau=0$) and is approximated by a power law for $\ell>1500$. The OV
contribution ($\Delta T^{OV}$) is conservatively taken to be $2.8\times
10^{-6}\sqrt{2\pi/\ell^2}$; i.e. $C_{\ell}^{OV}$  becomes important only at $\ell \gap 3000$.  For the noise calculation we assume that the
experimental beam is gaussian with width $\sigma_b\sim\theta_{FWHM}/2.3$,
where $\theta_{FWHM}$ is the beam full width at half maximum. Following Knox [23] we have that
$C_{\ell}^N=\exp(\ell^2\sigma_b^2)S$, where $S$ is the instrument sensitivity
i.e. the noise variance per pixel times the pixel solid angle in
steradiants.
For the noise contribution from many independent channels (as for Planck case
for example) we use  
$(C^N_{\ell})^{-1}=\sum_{\nu} (C^N_{ell}(\nu))^{-1}$.
  We consider two different experimental
settings. One with Planck specifications  the other with ACT for $\ell>200$
and MAP for $\ell<200$ specifications. 
ACT will map about $1/100$ of the sky with $2 \mu K$ per pixel
noise and experimental beam with  $\theta_{FWHM}=1.7'$.
Details about Planck specifications
can be found e.g., [7]. In practice, for the combination of ACT and MAP datasets,  useful signal can be extracted up to
$\ell \simeq 9000$ while for Planck up to $\ell \simeq 2000$.

\begin{figure}
\includegraphics[width=9cm]{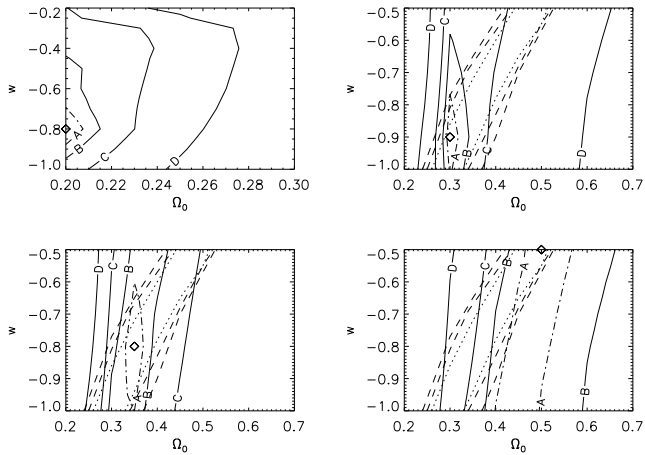}
\caption{Confidence levels in the $\Omega_0$-$w_Q$ plane. The dashed
and dotted contours show the degeneracy in the plane for the 2-year
MAP power spectrum data (see text for details). The other lines show
the expected confidence levels from the bispectrum analysis described
in the text applied to MAP and ACT data for different fiducial models
(indicated by the diamond). In particular solid lines show the
$68.3$\% (labelled by $B$), $90$\% (C) and $99$\% (D) confidence
region for $\Omega_0$ and $w_Q$ jointly considering only $\ell>13$ and the
small angle approximation. 
We estimated that by considering $\ell >2$ the constraints on
cosmological parameters  are improved as follows: the 68.3\% confidence level
region is indicated by the dot-dashed line labeled by $A$, the 90\%
confidence level correspond to the line labeled by $B$ and the $99\%$
correspond to the line labelled by $C$. Note that for low-$\Omega_0$
fiducial model the constraint are much more stringent than for
high-$\Omega_0$ model.}
\label{fig:fig1}
\end{figure}

\subsection{Breaking the degeneracy}

While observations of the microwave background fluctuations
are sensitive probes of cosmological parameters, there
are significant parameter degeneracies.  In a flat universe,
the position of the first acoustic peak depends primarily
on the angular diameter distance to the surface of last scatter.
In a universe with dark energy, this distance is a function 
of $\Omega_0$ and $w_Q$.  We have explored the degeneracy
by simulating a Monte Carlo Markov chain analysis
of the MAP 2 year data.  In the analysis, we have assumed
that the MAP data is limited by the statistical errors
and applied the Monte Carlo Markov Method developed in [24].  In our analysis,
we have fit the data with a seven parameter model
(power spectrum amplitude, power spectrum slope,
baryon density, matter density, Hubble constant, reionization
redshift and $w_Q$).

In Fig. \ref{fig:fig1} we show confidence levels in the $\Omega_0$-$w_Q$
plane for different fiducial models. 
The two dashed contours and the dotted contours show respectively the 99\%,
90\% and 68.3\% from the 2-year MAP power spectrum data. The other lines show the expected confidence levels from the bispectrum
analysis for MAP and ACT data.
Solid lines show the
$68.3$\% (labelled by $B$), $90$\% (C) and $99$\% (D) confidence region for $\Omega_0$ and $w_Q$
jointly. To obtain these contours the $\chi^2$ has been computed conservatively considering only
$\ell>12$ and the small angle approximation. However  the constraints on
cosmological parameters can be improved by considering all $\ell >2$. In this
case the 68.3\% confidence level region is indicated by the dot-dashed line labeled by $A$, the 90\%
confidence level correspond to the line labeled by $B$ and the $99\%$
correspond to the line labelled by $C$. Note that for low-$\Omega_0$
fiducial model the
constraint are much more stringent than for high-$\Omega_0$ model.

Similar constraints can be obtained from an experiment with the specifications
of the Planck satellite.

\section{Conclusions}

\begin{figure}
\includegraphics[width=9cm]{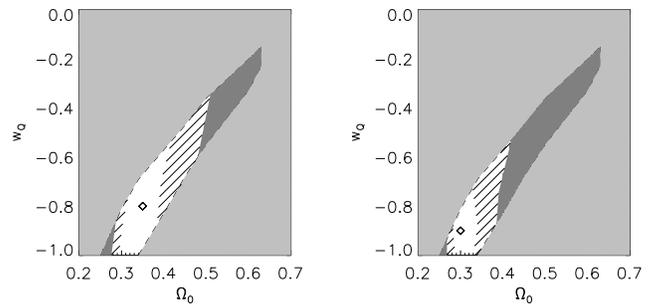}
\caption{90\% confidence level joint constraints on $\Omega_O$, $w_Q$
from MAP 2-year data and ACT for two fiducial models (indicated by the
diamond). Light gray shaded area is excluded by MAP power spectrum
analysis alone, while dark gray shaded area is excluded by MAP+ACT
small scale bispectrum considering only $\ell>12$. We extrapolate that
the area filled with pattern can be excluded by considering also $2<
\ell <12$ (see text for details). Similar constraints can be obtained from an
experiment with the specifications of the Planck mission.}
\label{fig:figconcl}
\end{figure}

We have computed the effect on the CMB bispectrum of the coupling between
Rees-Sciama, gravitational lensing, and primordial signal. This signal is
determined by the balance of two competing contributions along the line of sight: the decaying
gravitational potential fluctuations and the amplification due to linear
gravity. Both of these effects, and thus the bispectrum itself,  depend on $w_Q$ and $\Omega_0$. 
Since most of the bispectrum signal comes from the coupling of large scales
(low $\ell$) with small scales (large $\ell$) we have examined two
experimental settings that allow to accurately measure CMB temperature
fluctuations at low and high $\ell$'s: one is a combination of MAP 2-year data
with ACT CMB maps and the other has the specifications of the Planck surveyor.
As shown in Fig. (\ref{fig:figconcl}) we conclude that  we can realistically achieve an error of about 10\% on
$\Omega_0$ and $30\%$ on $w_Q$ at the 90\% joint  confidence level, by combining the constraints
from CMB power spectrum and bispectrum.

It is however important to bear some caveats in mind.
In general $w_Q$ might be time dependent. The CMB power spectrum will give
constraints on some `` weighted mean'' of $w_Q(z)$. The analysis presented here
constraints a {\it different} weighted mean of $w_Q(z)$, where most of the weight
comes from $z\sim 1$. This method has to be interpreted as a first  order
approximation to detect $w_Q \neq -1$.

We have also assumed that the CMB primordial signal is gaussian and that other
foregrounds contributions to the bispectrum (e.g., dust, point sources,
SZ effect) can be subtracted out.
While the SZ and point sources contributions can be accurately subtracted out
(e.g., [10]), dust should be negligible above the galactic plane and accurate
dust templates are available [25], the presence of a primordial
non-gaussian signal might invalidate the results.

\begin{acknowledgements}
We would like to thank Eiichiro Komatsu, Arthur Kosowski and  Chung-Pei Ma for
useful comments. 
We acknowledge the use of CMBFAST [22]. LV acknowledges the support of NASA grant NAG5-7154.
\end{acknowledgements}

\section{Appendix}

The derivation of Eq. (\ref{eq:smallangleql}) is conceptually similar to that of Spergel \&
Goldberg [15] for the Integrated Sachs-Wolfe effect, but is
complicated by the fact that the nonlinear evolution of the power
spectrum cannot be factorized in a time dependent and a scale dependent
functions.
We start from:
\ba
\langle\Theta^{*\ell_1}_{m_1} a_{\ell_2}^{m_2}\rangle=
-4 \left\langle\int d\hgamma_1 d\hgamma_2 dr\right.
\frac{r_*-r}{r_*r}\Phi^{NL}(r,\hgamma_1r)\nn
\times\left. \int d\tau \dot{\Phi}^{NL}(\tau,\hgamma_2\tau)
Y^{*m_1}_{\ell_1}(\hgamma_1)Y^{m_2}_{\ell_2}(\hgamma_2) \right\rangle
\ea

where the dot denotes $\partial/\partial \tau$.
Writing $\Phi$ in terms of its Fourier transform $\tilde{\Phi}$ and
expanding the exponential as $\exp(k\cdot \hgamma
r)=(4\pi)\sum_{\ell'm'}i^{\ell'}Y^{*m'}_{\ell'}(\hgamma_k)Y^{m'}_{\ell'}(\hgamma)j_{\ell'}(kr)$,
we obtain:
\ba
\langle\Theta^{*\ell_1}_{m_1} a_{\ell_2}^{m_2}\rangle \!\!\!&\!\! \simeq &\!\!\!\!-2 (4\pi)^2\!\!\int\!
d\hgamma_1 d\hgamma_2 dr\frac{r_*-r}{r_*r}\!\int\! d\tau
\frac{d^3k}{(2\pi)^3} i^{\ell'+\ell''} \nn
 &\!\!\!\!\!\! \times &\!\!\!\!\! \dot{P}_{\Phi}(k;\tau,r) i^{\ell'+\ell''}j_{\ell'}(kr) Y^{*m'}_{\ell'}(\hgamma_k)Y^{m'}_{\ell'}(\hgamma_1)\nn
& \!\!\!\!\!\! \times &\!\!\!\!\!\! Y^{*m''}_{\ell''}(\hgamma_k)Y^{m''}_{\ell''}(\hgamma_2)Y^{*m_1}_{\ell_1}(\hgamma_1)Y^{m_2}_{\ell_2}(\hgamma_2)
\ea
where $P_{\Phi}(k;\tau,r)$ is defined through 
\be
\langle \dot{\tilde{\Phi}}(k,\tau)
\tilde{\Phi}(k',r)\rangle=2(2\pi)^3\dot{P}_{\Phi}(k;\tau,r)\delta^D({\bf k}+{\bf k}')
\ee
and $\delta^D$ denotes the Dirac delta function.
In principle (21) has an extra term which vanishes at high $\ell$.
Using the fact that $d^3k=k^3 dk d\hgamma_k$ and the orthogonality relations of
spherical harmonics, we obtain $\int d\hgamma_1 d\hgamma_2 d\hgamma_k$ yields
$-\delta^K_{\ell'\ell_1}\delta^K_{m_1 m'}
\delta^K_{\ell''\ell_2}\delta^K_{m_2 m''}\delta^K_{\ell''\ell'}\delta^K_{m''m'}$ , where $\delta^K$ denotes the
Kronecker delta. Finally using the approximation:
\be
\int dk k^2 f(k)j_{\ell'}(kr)k_{\ell''}(k\tau)\simeq
f(\ell'/r)\pi/2r^2\delta(r-\tau)
\ee
and performing the remaining integral in $d\tau$, we obtain Eq. \ref{eq:smallangleql}.

\end{document}